\documentclass[aps,prx,twocolumn,superscriptaddress,showpacs,floatfix,longbibliography,dvipsnames]{revtex4-2}
\usepackage{amsmath,amssymb,amsfonts,graphics,epsfig,epstopdf,color,verbatim,tabularx,bm,multirow,appendix,hyperref}
\pdfoutput=1
\usepackage[normalem]{ulem}
\usepackage{lmodern}
\usepackage{ytableau}
\usepackage{pifont}
\usepackage{color}
\usepackage{bm}
\usepackage{wasysym}
\usepackage{dsfont}
\DeclareMathOperator{\Tr}{Tr}

\hypersetup{
	colorlinks,
	linkcolor={BrickRed},
	citecolor={BrickRed},
	urlcolor={BrickRed},
	pdftitle={},
	pdfauthor={},
	pdfstartview={FitH}
}

\def\bq{{\mathbf{q}}}

\def\bn{{\mathbf{n}}}
\def\bsigma{{\mathbf{\sigma}}}

\begin{document}
	
\title{Bilinear correlations in Fluctuating Gaussian States with anti-unitary symmetries}

\author{Xu Zhang}
\affiliation{Department of Physics and Astronomy, Ghent University, Krijgslaan 299, 9000 Gent, Belgium}

\begin{abstract}
We study bilinear order-parameter correlations in a general class of sign problem-free systems that are described by what we call Fluctuating Gaussian States (FGS) with anti-unitary symmetries, where the weight of each Gaussian measurement in the space-time path integral is positive-definite. Supported by Monte Carlo simulations on fluctuating Gaussian fermionic and bosonic examples, we argue that such systems generally exhibit two types of broken-symmetry phases: a \emph{saddle point phase} and a \emph{non-local contraction phase}, with the competition between the two determined by the stability of the fluctuation saddle point of FGS. Due to the Fermi liquid instability in fermionic FGS with anti-unitary symmetries, we also discuss the possibility of representing fermionic FGS with a particular Projected Entangled-Pair State (PEPS) ansatz, which we argue a certain construction provides an efficient description for the saddle point phase, but not for the non-local contraction phase. Finally, we discuss the potential implication of our result in non-equilibrium systems and stabilizing a target bilinear order.
\end{abstract}
\date{\today}
\maketitle

\section{Introduction}
Variational Matrix Product States (MPS), after many years of development, have become one of the most powerful numerical methods for simulating one-dimensional quantum systems~\cite{schollwock2011density,cirac2021matrix}. The Projected Entangled-Pair States (PEPS) ansatz, a generalization of MPS to higher dimensions as proposed by Verstraete and Cirac~\cite{verstraete2004renormalization}, has also shown its potential in representing different quantum phases with area-law entanglement ~\cite{cirac2021matrix}. Quantum Monte Carlo (QMC), on the other hand, is not limited by the system dimension~\cite{Assaad2008,gull2011continuous}, but requires the probability weights for sampling to be positive and hence is faced with the \emph{sign problem}~\cite{loh1990sign,PAN2024879}. Following Wu and Zhang~\cite{wusufficient2005}, a broad variety of sufficient conditions was established to guarantee the absence of a sign problem in fermion QMC~\cite{Lisolving2015,weimajorana2016,Limajorana2016,lisign2019,wang2015split,wei2024semigroup,chandrasekharan2010fermion,huffman2014solution,huffman2016solution}, and Refs.~\cite{Ringel2017,Ringel2020,Ringel2020_2,Ellison2021} also made general connections between the sign problem and the restriction for representing topological quantum phases. Nevertheless, the precise constraints on physical properties implied by those sufficient conditions to remove the sign problem (like the entanglement constraint in the PEPS ansatz) are not yet completely understood, and only recently some progress towards answering this question has been made. Grossman and Berg have shown that a robust Fermi liquid cannot exist in sign-problem-free models~\cite{grossmanrobust2023}, and the author together with Bultinck has shown that a Kramers (i.e. $\mathcal{T}^2=-1$) anti-unitary symmetry, which removes the sign problem, imposes stringent constraints on correlation functions of fermion bilinears~\cite{zhang2025constraints,zhang2025bilayer}.

\begin{figure}[htp]
\centering
\includegraphics[width=1.0\columnwidth]{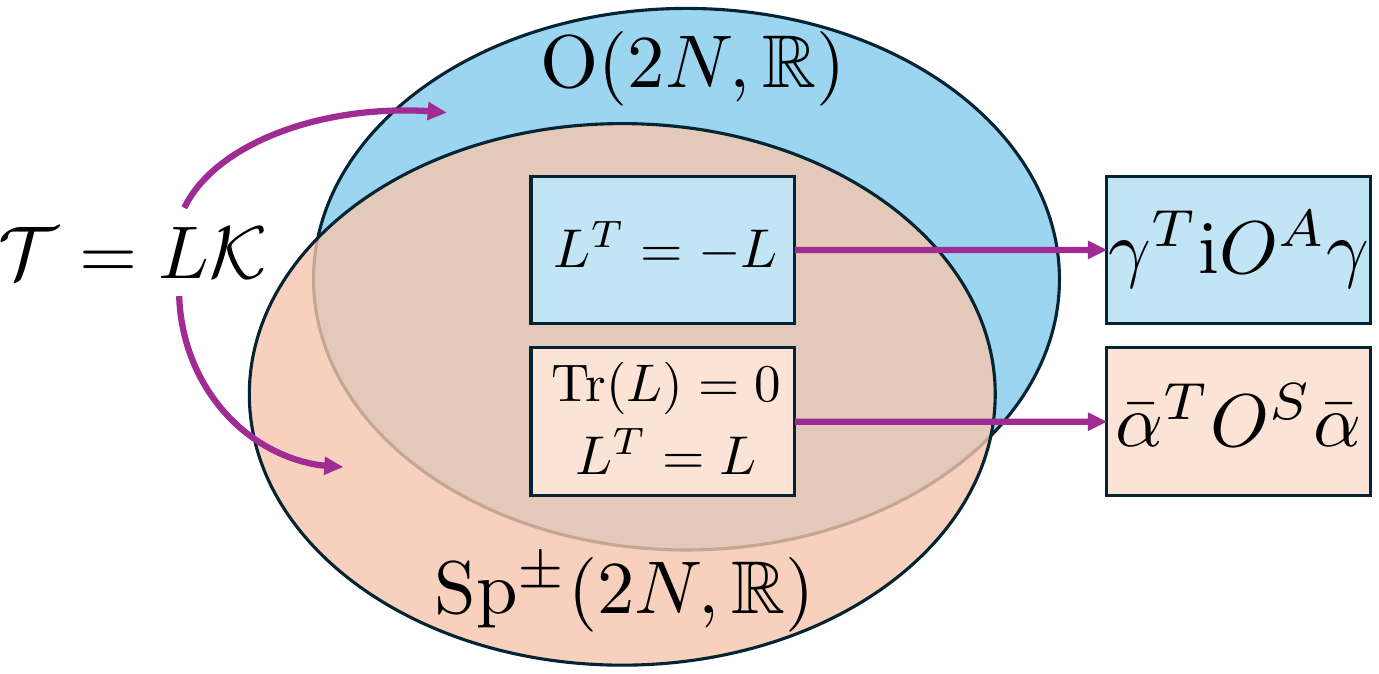}
\caption{The schematic relationship between anti-unitary symmetry and order parameter set for Majorana fermion (up blue) and phase-space boson (down orange).}
\label{symo}
\end{figure}

In this paper, we expand our understanding of the physical implications of anti-unitary symmetries by firstly introducing the notion of \textit{Fluctuating Gaussian States} (FGS) for Majorana fermions (phase-space bosons) with anti-symmetric (traceless-symmetric) anti-unitary symmetries $\mathcal{T}=L\mathcal{K}$. As shown in Fig.~\ref{symo}, canonical basis transformation requires a fermionic/bosonic FGS with $N$ orbitals has $L\in\text{O}(2N,\mathbb{R})/\text{Sp}^\pm(2N,\mathbb{R})$, where we define the set composed from symplectic and anti-symplectic matrices as the group $\text{Sp}^\pm(2N,\mathbb{R})\equiv\{L\in\text{GL}(2N,\mathbb{R})|L^T\Omega L=\pm\Omega\}$. By showing any anti-symmetric (traceless-symmetric) Hermitian matrix within bilinear observables can be decomposed into matrices which preserve both orthogonal and symplectic or anti-symplectic condition, we define a complete order parameter set $\{\gamma^T\mathrm{i}O^A\gamma\}$ ($\{\bar{\alpha}^TO^S\bar{\alpha}\}$) for Majorana fermions (phase-space bosons). Correlation functions of those order parameter bilinears in FGS have two types of contributions: one from direct Wick contraction, and one from non-local (i.e. exchange) Wick contraction. We prove that: \emph{If FGS have an anti-unitary symmetry $\mathcal{T}=L\mathcal{K}$ where matrix $L$ belongs to a bilinear in the order parameter set, non-local contraction of the corresponding order parameter will be maximal compared with those from other elements in the set}.

The main text of this paper is organized as follows. In Sec.~\ref{s2} and Sec.~\ref{s3}, we give detailed definition of FGS, anti-unitary symmetry and order parameter set. In Sec.~\ref{s4}, we prove the main conclusion mentioned above and further discuss the generic competition in FGS between phases with different long-range orders, which show up in either the direct or exchange contributions to the correlation functions. In particular, we argue that the direct contribution dominates when a mean-field saddle-point analysis is applicable for the FGS, and the non-local contraction dominates when including the fluctuation around a saddle-point is essential. After introducing an engineered way of tuning the fluctuations in Sec.~\ref{s5}, we numerically confirm this statement with fermionic and bosonic FGS examples in Sec.~\ref{s6} and Sec.~\ref{s7}. Since a stable Fermi liquid is ruled out for fermionic FGS, in Sec.~\ref{s8} we utilize our result of correlation functions and discuss the possibility of representing such FGS with a particular PEPS ansatz construction. Finally, we conclude in Sec.~\ref{s9} and discuss the potential implication for non-equilibrium systems and stabilizing a target bilinear order.

\section{Fluctuating Gaussian states}\label{s2}
We first introduce our definition of Fluctuating Gaussian States. The density matrix of a Gaussian state~\cite{lucas2021bosonic,lucas2020geometry,braunstein2005quantum,bravyi2005lagrangian,WANG20071,adesso2014continuous} is defined as
\begin{equation}
\rho=e^{-\bar{c}^T Q \bar{c}-q},
\end{equation}
where $\bar{c}\equiv c-z$, $c_{i}=\frac{1}{\sqrt{2}}(a^\dagger_{i}+a_{i}),c_{2i}=\frac{\mathrm{i}}{\sqrt{2}}(a^\dagger_i-a_i)$ are Hermitian operators representing either Majorana fermions ($c\rightarrow\gamma$) or phase-space bosons ($c\rightarrow\alpha$) at orbital $i$ and we assume the total orbital number is $N$, $z_{i}=\left\langle c_{i}\right\rangle \equiv \Tr(\rho c_{i})$ is the one-point function which is zero for fermions, $q$ is a normalization constant (to impose $\Tr(\rho)=1$), and $Q$ is an anti-symmetric (symmetric) $2N \times 2N$-dimensional matrix for fermions (bosons). Note that we do not require $\rho$ to be Hermitian, since this character is not promised in general FGS. The important property of Gaussian states is one can use Wick's theorem to compute many-point correlation functions from two-point correlation functions, e.g. $\left\langle \bar{c}_i \bar{c}_j \bar{c}_k \bar{c}_l \right\rangle = \left\langle \bar{c}_i \bar{c}_j\right\rangle \left\langle \bar{c}_k \bar{c}_l\right\rangle + \left\langle \bar{c}_i \bar{c}_l\right\rangle \left\langle \bar{c}_j \bar{c}_k\right\rangle \mp \left\langle \bar{c}_i \bar{c}_k\right\rangle \left\langle \bar{c}_j \bar{c}_l \right\rangle$,
where $\mp$ applies for fermionic/bosonic Gaussian states.

The core spirit of the Green's function based QMC method is to rewrite the observables of an interacting system as a weighted average of Gaussian measurements via a space-time local bosonic field $\left\{\phi\right\} $. For a fixed configuration, an effective density matrix $\rho_{\phi}\equiv\prod_{\tau=1}^{L_{\tau}}B_\phi(\tau)$ is composed from a product of Gaussian states from each imaginary time slice after a Suzuki–Trotter decomposition, the result of which is still Gaussian. For example, the two-point correlation of Majorana fermion is $\left\langle \gamma_i(\tau) \gamma_j(\tau') \right\rangle \equiv\frac{\int [\mathcal{D} \phi] e^{-S(\left\{\phi\right\})} \left\langle \gamma_i(\tau) \gamma_j(\tau') \right\rangle_{\phi}}{\int [\mathcal{D} \phi'] e^{-S(\left\{\phi'\right\})}}$,
where we use $\tau,\tau'$ to label time coordinate of the operator, and $\langle\cdots\rangle_\phi\equiv \Tr(\mathcal{P} \{\rho_\phi \cdots\})$ is an expectation value of the Gaussian state with (imaginary time) path-ordering within $\mathcal{P}\{\cdots\}$.

Inspired by the QMC formula, if the measurement in a system can be written as
\begin{equation}
    \left\langle \cdots \right\rangle \equiv\frac{\int [\mathcal{D} \phi] e^{-S(\left\{\phi\right\})} \left\langle \cdots \right\rangle_{\phi}}{\int [\mathcal{D} \phi'] e^{-S(\left\{\phi'\right\})}},
\end{equation}
where $S(\left\{\phi\right\})$ is real for any configuration and the coupling between the field $\phi$ and the Gaussian fermion/boson operators is space-time local and homogeneous, we call such a system \textit{Fluctuating Gaussian States} (FGS). It is straightforward to see FGS can be simulated by QMC without sign problem, while sign-problem-free models have an additional requirement that $Z=\int [\mathcal{D} \phi] e^{-S(\left\{\phi\right\})}$ should define a physical partition function.

In fermion QMC, the existence of two anti-unitary symmetries $\mathcal{T}=O\mathcal{K}$, where $O$ is a real orthogonal transformation on the Majorana fermions and $\mathcal{K}$ is complex conjugation, is typically used to prove that $S(\left\{\phi\right\})$ is real~\cite{wusufficient2005,Lisolving2015,Limajorana2016,weimajorana2016,lisign2019}. In a previous study~\cite{zhang2025constraints,zhang2025bilayer}, we notice that a Kramers anti-unitary symmetry where $O$ is product of Pauli matrices can also indicate the maximal bilinear correlation function in the non-local contraction term. Below, we will generalize this result to fermionic (bosonic) FGS with anti-unitary symmetry $\mathcal{T}=L\mathcal{K}$, where $L$ is real orthogonal (symplectic or anti-symplectic) matrix.

\section{Bilinear observables under anti-unitary symmetries}\label{s3}
When we say the FGS has an anti-unitary symmetry, we mean the Gaussian state at every $\tau$ has
\begin{equation}
B_\phi(\tau)=B_\phi^*(\tau,\bar{c}\rightarrow L\bar{c})
\end{equation}
where $L\in\text{O}(2N,\mathbb{R})/\text{Sp}^\pm(2N,\mathbb{R})$ for fermionic/bosonic FGS and $S(\left\{\phi\right\})$ is invariant~\cite{antiunitary}. With this anti-unitary symmetry, the bilinear observable defined from a Hermitian matrix $M$ has the property
\begin{equation}
\label{mms}
\left\langle \bar{c}^{T}(\tau) M^T \bar{c}(\tau') \right\rangle_\phi = \left\langle \bar{c}^{T}(\tau) L^T M L \bar{c}(\tau') \right\rangle_\phi^*,
\end{equation}
where we have used linear/anti-linear canonical basis transformation for real orthogonal or real symplectic/anti-symplectic matrix $L$ (see Appendix.~\ref{ct} for details). Now considering a basis $\{T^a\}_{a=1}^{4N^2}$ of $2N\times 2N$-dimensional Hermitian matrices normalized according to $\Tr(T^aT^b)=\delta^{ab}$ (i.e. generators of the group U($2N$)~\cite{Georgi2000Lie}), every $2N\times 2N$-dimensional matrix can be written as a \emph{complex} linear combination $u^aT^a$, where here and below summation over the repeated index is assumed. Specializing to $LG(\tau',\tau)L^T$, with the two-point Green's function $G_{ij}(\tau',\tau)\equiv\left\langle \bar{c}_j(\tau) \bar{c}_i(\tau')\right\rangle_\phi$, we have the coefficient
\begin{equation}
    \label{vT}
    v^a=\Tr(LG(\tau',\tau)L^T T^a)=\left\langle \bar{c}^{T}(\tau) L^T T^a L \bar{c}(\tau') \right\rangle_\phi.
\end{equation}
Since $T^a$ is Hermitian, we can use Eq.~\eqref{mms} to show that the Green's function satisfies
\begin{equation}
    \label{gf}
    v^aT^a\equiv LG(\tau',\tau)L^T=G^*(\tau',\tau).
\end{equation}
Since the equation holds for every configuration, it is also a property of the two-point Green's function of FGS with anti-unitary symmetry. We will use Eq.~\eqref{gf} to derive one of our main results regarding bilinear correlation functions in the next section. Before that, we first define the order parameter set.

Starting from a general Hermitian bilinear observable $\bar{c}^T M \bar{c}$, one can use the basis $\{T^a\}_{a=1}^{4N^2}$ to expand the Hermitian matrix $M$ as $M=m^a T^a$ with \emph{real} coefficient $m^a = \Tr(M T^a)$. Except for the identity which transforms trivially, the other traceless matrices have $T^a=\Re(T^a)+\mathrm{i}\Im(T^a)$, where the imaginary (real) part is real anti-symmetric (traceless-symmetric) matrix corresponding to fermionic (bosonic) observables. We separate the $2\times2$ Majorana space/phase space out with $\kappa^{b}\in\{\mathds{1},\sigma^x,\mathrm{i}\sigma^y,\sigma^z\}$, where $\bsigma$ are Pauli matrices. By noticing an anti-symmetric (symmetric) matrix in its block-diagonalized standard form can further be decomposed into linear combination of real \emph{orthogonal} anti-symmetric (symmetric) matrices, the matrix in the left space after separation can be decomposed accordingly. i.e.,
\begin{eqnarray}
    \Im(T^a)&=& C^{a;b,c}\kappa^b\otimes O^c,\\
    \Re(T^a)&=& D^{a;b,c}\kappa^b\otimes O^c.\nonumber
\end{eqnarray}
The reason we treat Majorana space/phase space specially is one can check $\{\mathds{1},\mathrm{i}\sigma^y\}$ and $\{\sigma^x,\sigma^z\}$ correspond to particle-number preserving and breaking orders by transforming $\bar{c} \kappa\otimes O \bar{c}$ back into Fock space, and $\kappa^b\in\{\mathds{1},\mathrm{i}\sigma^y\}/\{\sigma^x,\sigma^z\}$ also corresponds to symplectic/anti-symplectic decomposition separately. Those matrices form a subset of $\text{O}(2N,\mathbb{R})\cap\text{Sp}^\pm(2N,\mathbb{R})$ in anti-unitary symmetries we defined above, thus when such an anti-unitary symmetry exists we can always find the corresponding observable component.

As shown in Fig.~\ref{symo}, we call the decomposed real orthogonal anti-symmetric matrices $O^A$ (traceless-symmetric matrices $O^S$) so that $M=w^a [\mathrm{i}O^A]^a$ ($M=w^a [O^S]^a$) represent general fermionic (bosonic) observables which transform non-trivially. By simply taking $\left\{\gamma^{T}\mathrm{i} O^A \gamma\right\}$ ($\left\{\bar{\alpha}^{T} O^S \bar{\alpha}\right\}$), we find a complete order parameter set for all bilinear observables (i.e., when a symmetry which can be detected with local bilinear observables is broken, the measurement of elements transforming non-trivially will develop a finite density). Within a highly symmetric phase (i.e. $\rho_\phi\rightarrow1$), those order parameters transform into each other with canonical basis transformation and there is no preference between measurements of them, so that we can compare them evenly for symmetry breaking. The comparison of bilinear correlations between order parameters in the next section is based on this set. And we use $O$ and $L$ to distinguish the matrix in order parameters and anti-unitary symmetries from now on, even though the utilized $L$ is also orthogonal.

\section{Bilinear Correlation under anti-unitary symmetries}\label{s4}
When $\phi$ is fixed, the correlation function of bilinear order parameters can be expressed with Wick's theorem as
\begin{eqnarray}
\label{lala}
&&\left\langle [\bar{\alpha}^T(\tau) O^S \bar{\alpha}(\tau)] [\bar{\alpha}^T(\tau') O^S \bar{\alpha}(\tau')] \right\rangle_\phi \\
&=& \left\langle \bar{\alpha}^T(\tau) O^S \bar{\alpha}(\tau) \right\rangle_\phi \left\langle \bar{\alpha}^T(\tau') O^S \bar{\alpha}(\tau') \right\rangle_\phi \nonumber\\
&+& O^S_{ba}O^S_{cd}\left[G_{da}(\tau',\tau) G_{cb}(\tau',\tau) + G_{ca}(\tau',\tau) G_{db}(\tau',\tau) \right], \nonumber\\\nonumber\\
&&\left\langle [\gamma^{T}(\tau)\mathrm{i} O^A\gamma(\tau)] [\gamma^{T}(\tau')\mathrm{i} O^A\gamma(\tau')] \right\rangle_\phi \nonumber\\
&=& \left\langle \gamma^{T}(\tau)\mathrm{i} O^A\gamma(\tau) \right\rangle_\phi \left\langle \gamma^{T}(\tau')\mathrm{i} O^A\gamma(\tau') \right\rangle_\phi \nonumber\\
&+& O^A_{ba}O^A_{cd}\left[G_{da}(\tau',\tau) G_{cb}(\tau',\tau) - G_{ca}(\tau',\tau) G_{db}(\tau',\tau) \right]. \nonumber
\end{eqnarray}
Note the element product of $O$ can be decomposed with the basis $\{T^a\}_{a=1}^{4N^2}$ as
\begin{equation}
O_{ba} O_{cd} = \left[O^T T^g O\right]_{ad} \left[T^g\right]_{cb}, \label{llt}
\end{equation}
This can be easily proved by noticing that the right part of the equation can be reduced as
\begin{eqnarray}
&&\left[O^T T^g O\right]_{ad} \left[T^g\right]_{cb} \\
&=& O_{ma} O_{nd} \left[T^g\right]_{mn} \left[T^g\right]_{cb} \nonumber\\
&=& O_{ma} O_{nd} \delta_{mb}\delta_{cn} \nonumber\\
&=& O_{ba} O_{cd}. \nonumber
\end{eqnarray}
Here in the third line we have used the completeness relation of the basis, i.e. $\left[T^g\right]_{mn} \left[T^g\right]_{cb}=\delta_{mb}\delta_{cn}$. With Eq.~\eqref{llt} and Eq.~\eqref{gf}, we can rearrange the labels of real orthogonal matrices in Eq.~\eqref{lala} and express the correlation function as
\begin{eqnarray}
\label{CF}
&&\left\langle [\bar{\alpha}^T(\tau) O^S \bar{\alpha}(\tau)] [\bar{\alpha}^T(\tau') O^S \bar{\alpha}(\tau')] \right\rangle_\phi \\
&=& \left\langle \bar{\alpha}^T(\tau) O^S \bar{\alpha}(\tau) \right\rangle_\phi \left\langle \bar{\alpha}^T(\tau') O^S \bar{\alpha}(\tau') \right\rangle_\phi \nonumber\\ 
&+& 2\Tr[(O^S)^T T^g O^S G(\tau',\tau)] \Tr[ L^T T^g L G(\tau',\tau)]^*, \nonumber\\\nonumber\\
&&\left\langle [\gamma^{T}(\tau)\mathrm{i} O^A\gamma(\tau)] [\gamma^{T}(\tau')\mathrm{i} O^A\gamma(\tau')] \right\rangle_\phi \nonumber\\
&=& \left\langle \gamma^{T}(\tau)\mathrm{i} O^A\gamma(\tau) \right\rangle_\phi \left\langle \gamma^{T}(\tau')\mathrm{i} O^A\gamma(\tau') \right\rangle_\phi \nonumber\\
&+& 2\Tr[(O^A)^T T^g O^A G(\tau',\tau)] \Tr[ L^T T^g L G(\tau',\tau)]^*. \nonumber
\end{eqnarray}
Note in Eq.~\eqref{CF}, the fermion/boson correlations start sharing the same structure, i.e. the first direct contraction term and the second non-local contraction term. Following from $v^aT^a=O G(\tau',\tau) O^T$ in Eq.~\eqref{gf} and the hermiticity of $T^a$, the norm squared $v^a(v^a)^*$ is independent of $O$ and $T^a$, i.e.
\begin{equation}
v^a (v^a)^*=\Tr(v^a T^a (v^b)^* T^b) = \left\lVert G(\tau',\tau) \right\rVert_F^2,
\end{equation}
where $\left\lVert G\right\rVert_F\equiv\sqrt{\Tr(G G^\dagger)}$ is the Frobenius norm of the matrix $G$. From the Cauchy-Schwarz inequality of vector $v^a$, we have
\begin{equation}
\left\lvert \Tr[O^T T^g O G(\tau',\tau)] \Tr[L^T T^g L G(\tau',\tau)]^* \right\rvert \leq \left\lVert G(\tau',\tau)\right\rVert_F^2,
\end{equation}
where the equality holds iff $O=L$. This inequality builds up our statement: For anti-unitary symmetric fermionic (bosonic) FGS with anti-symmetric (traceless-symmetric) $L\in\text{O}(2N,\mathbb{R})\cap\text{Sp}^\pm(2N,\mathbb{R})$, one can always use $L$ to define a bilinear correlation, whose non-local contraction is maximal compared with others from a complete order parameter set. Now we focus on the equal-time correlation to discuss the phase diagram of spontaneous symmetry breaking from bilinear orders.

\begin{figure}[htp]
\centering
\includegraphics[width=1.0\columnwidth]{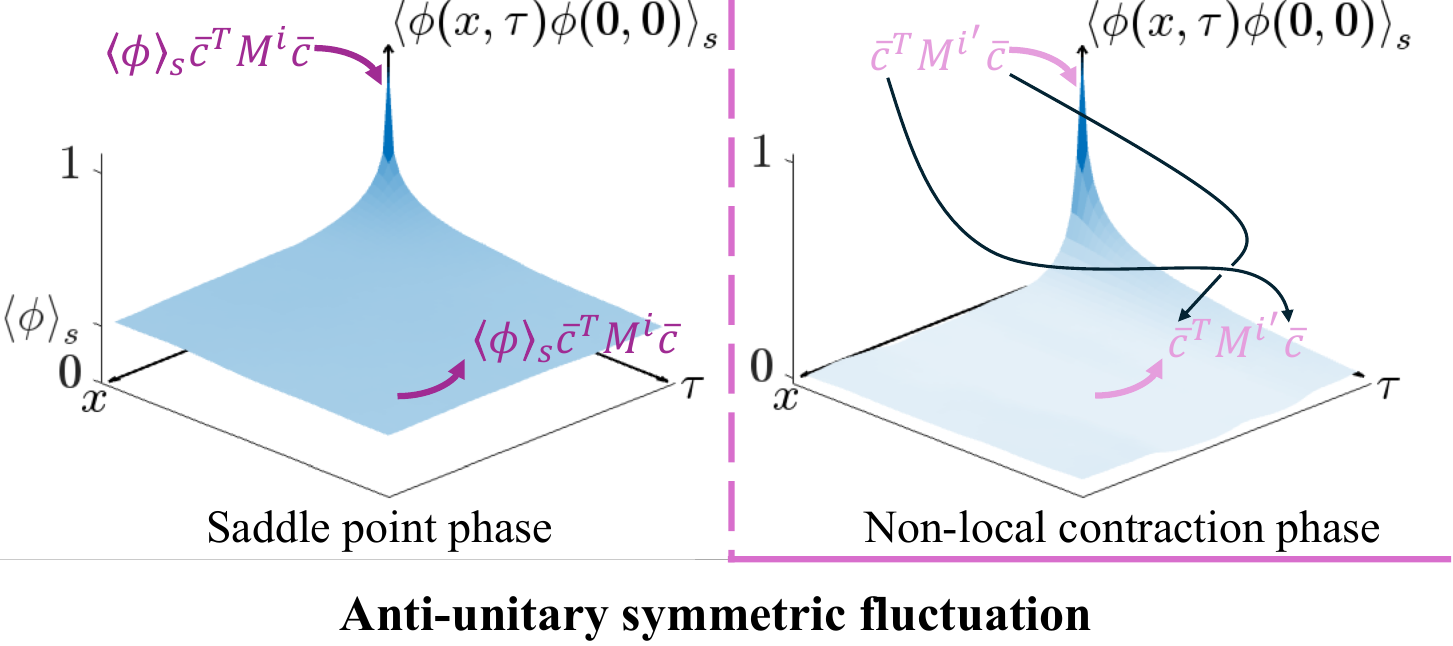}
\caption{The schematic phase diagram between two ordered phases of FGS with anti-unitary symmetry. The left/right part indicates the bilinear order selected by the saddle point of $S(\left\{\phi\right\})$ and the anti-unitary symmetry, respectively.}
\label{phd}
\end{figure}

To enter an ordered phase by spontaneously breaking a symmetry of FGS, the equal-time correlation function defined in Eq.~\eqref{CF} should scale with system size (i.e., $\sim\mathcal{O}(N^2)$) and has a finite density in the thermodynamic limit. One straightforward way is $S(\left\{\phi\right\})$ has several stable saddle points for the $\phi$ field. As shown in the left part of Fig.~\ref{phd}, at each saddle point the normalized space-time correlation $\left\langle \phi(x,\tau)\phi(0,0)\right\rangle_s$ decays from 1 to a constant $\left\langle \phi\right\rangle_s$. Assuming that the local and homogenous coupling has the form $\phi \bar{c}^{T} M^i \bar{c}$, the long-range order of $\phi$ will in turn drive the long-range order of $\bar{c}^{T} M^i \bar{c}$. The contribution from different symmetry-related saddle points cancels so that FGS are symmetric, while the square gives finite contribution $\sim\mathcal{O}(N^2)$, indicating spontaneous symmetry breaking. This corresponds to the mean-field order formed by the first term in Eq.~\eqref{CF}, and we call it \emph{saddle point phase} in Fig.~\ref{phd}.

The only other way to establish long-range order is from the second term in Eq.~\eqref{CF}, where as shown above the $O=L$ order is maximal, and requires at least $\left\lVert G\right\rVert_F^2=\sum_{i,j}\left\lvert G_{ij}\right\rvert^2\sim\mathcal{O}(N^2)$. We argue this order for fermions or uncondensed bosons cannot be realized close to any stable saddle point of $S(\left\{\phi\right\})$ whose space-time correlation decays exponentially. To see this, consider the expansion of $S(\left\{\phi\right\})$ around the saddle point $S(\left\{\phi\right\})\approx S(\left\{\phi_0\right\})+\delta\phi^{T} K \delta\phi$. The measurement of $\left\lvert G_{ij}\right\rvert^2$ is then
\begin{equation}
    \left\langle \left\lvert G_{ij}\right\rvert^2\right\rangle = \frac{\int [\mathcal{D} \delta\phi] e^{-\delta\phi^{T} K \delta\phi} |G_{ij}^{\phi_0 + \delta \phi}|^2}{\int [\mathcal{D} \delta\phi'] e^{-(\delta\phi')^{T} K \delta\phi'}}.
\end{equation}
Except for the zeroth order of $\delta\phi$ in $|G_{ij}^{\phi_0 + \delta \phi}|^2$ which is the measurement of a Gaussian state $|G_{ij}^{\phi_0}|^2$, the higher order contributions of $\delta \phi$ are exponentially small and hence can be ignored when the distance $r_{ij}$ becomes large since $K^{-1}_{ij}\sim e^{-r_{ij}/\xi}$. For a fermionic Gaussian state, the zeroth order at most has a power-law correlation, which is not enough to satisfy $\sum_{i,j}\left\lvert G_{ij}\right\rvert^2\sim\mathcal{O}(N^2)$. While for a bosonic Gaussian state, this finite-density requirement can be satisfied by either a trivial condensation, or similar to the fermionic case, by staying away from any stable saddle point. Thus, we expect that a power-law decay of $\left\langle \phi(x,\tau)\phi(0,0)\right\rangle_s$ at a critical saddle point can give rise to a situation where the second term in Eq.~\eqref{CF}, obtained from the non-local contraction, establishes the long-range order of $\bar{c}^{T}M^{i'}\bar{c}$, which as explained above is selected by the anti-unitary symmetry. And we name it the \emph{non-local contraction phase} as schematically represented in the right part of Fig.~\ref{phd}. To confirm this expectation, in the next sections we engineer critical fluctuations of $\phi$ as the leading order approximation of the expansion of $S(\left\{\phi\right\})$ around a critical saddle point and numerically study two examples.

\section{Engineered critical fluctuations}\label{s5}
In this section, we discuss how to generate a random scalar field satisfying a certain critical correlation. By coupling the critical fluctuating scalar field with fermionic/bosonic Gaussian states and tuning the critical exponents, we will show the competition between two different order phases in the next two sections.

\begin{figure}[htp]
\centering
\includegraphics[width=1.0\columnwidth]{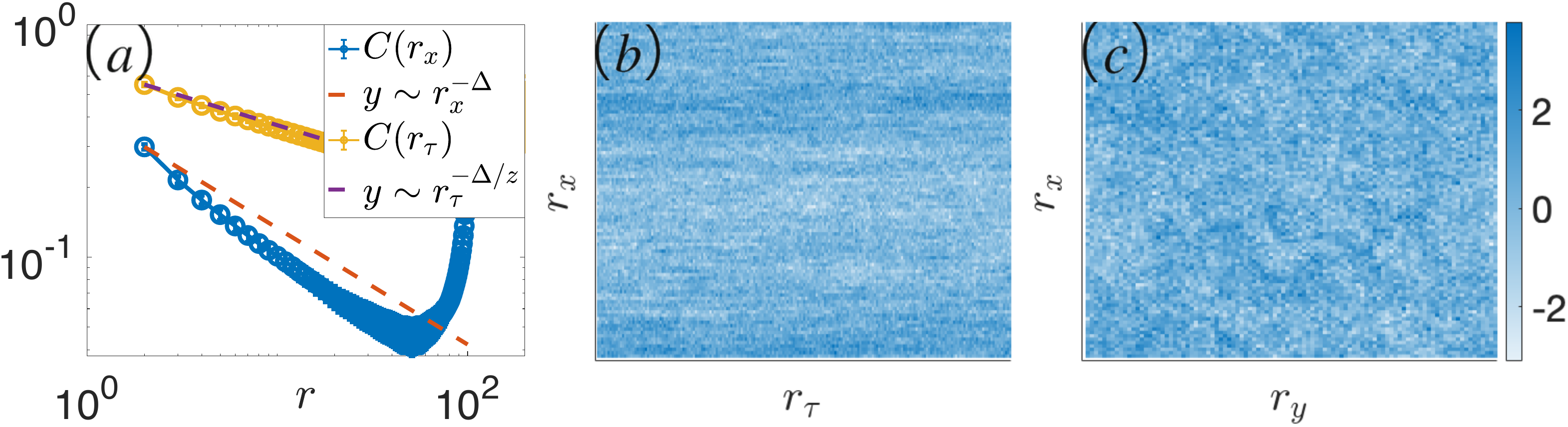}
\caption{An example of engineering scalar field $\phi$ for $\Delta=0.5,z=2$. (a) Correlation decay along $r_x$ and $r_\tau$ directions from 100 samples for size $L=100,L_{\tau}=200$ with $\Delta\tau=0.05$ are compared with the target power-law behaviors. (b,c) show a specific configuration cut with fixed $r_y$ and $r_\tau$ respectively.}
\label{Crall}
\end{figure}

Firstly, we assume the distance from $(r_x,r_y,r_\tau)$ to the origin point in the 2+1D space-time is defined as
\begin{equation}
    r = \sqrt{r_x^2+r_y^2+(r_\tau \Delta\tau)^{2/z}},
\end{equation}
where $r_x,r_y,r_\tau$ are integer coordinates of the discretized lattice, $\Delta\tau$ is the imaginary time step, and $z$ is the dynamical critical exponent. Assume the target correlation of a scalar field is defined as
\begin{equation}
    C(r)\equiv\langle \phi(r) \phi(0) \rangle = \frac{1}{(r/r_0+1)^{\Delta}},
\end{equation}
where $\Delta=d-2+\eta$ indicates the anomalous dimension $\eta$ in dimension $d=2$ and this correlation is normalized to 1 at $r=0$ with a short-range cutoff $r_0$ to control the window reaching the power-law decay (we set $r_0=0.1$ in our simulation). Such a scalar field $\phi$ could be realized by coupling the Fourier transformation of a random scalar field $\tilde{\phi}$ generated from a standard Gaussian distribution and the Fourier transformation of $C(r)$ as
\begin{equation}
    \phi(r) = \mathcal{F}^{-1} \left[\sqrt{\mathcal{F}\left[C(r)\right]} \mathcal{F}\left[\tilde{\phi}\right] \right].
\end{equation}
One can check the correlation is realized correctly by
\begin{eqnarray}
    \langle \phi(r) \phi(0) \rangle &=& \langle\mathcal{F}^{-1} \left[\mathcal{F}\left[\phi\right] \mathcal{F}\left[\phi\right]^* \right]\rangle \\
    &=&\mathcal{F}^{-1} \left[\mathcal{F}\left[C(r)\right] \langle \mathcal{F}\left[\tilde{\phi}\right] \mathcal{F}\left[\tilde{\phi}\right]^* \rangle \right] \nonumber\\
    &=&C(r), \nonumber
\end{eqnarray}
where in the last step we have used the fact that $\tilde{\phi}$ is generated from standard Gaussian distribution and thus satisfies $\langle \mathcal{F}\left[\tilde{\phi}\right] \mathcal{F}\left[\tilde{\phi}\right]^* \rangle=1$. One example for $\Delta=0.5,z=2$ is shown in Fig.~\ref{Crall} for size $L=100,L_{\tau}=200$ with $\Delta\tau=0.05$. In panel (a), $C(r)$ along $r_x$ and $r_\tau$ computed from 100 samples are compared with the target power-law decay indicated by dashed lines. A specific configuration cut at fixed $r_y$ and $r_\tau$ are shown in panels (b) and (c) respectively.

\section{Simulation of fermionic FGS with engineered critical fluctuations}\label{s6}
In this section, we consider a specific fermionic FGS example, where the unnormalized  $\rho_\phi$ is
\begin{equation}
\rho_\phi=\prod_{\tau=1}^{L_\tau}e^{\Delta\tau (\sum_{i,j} t_{i,j} \gamma^T_i \kappa^y l^0 \sigma^0 \gamma_j + g \sum_{i}(-1)^{i_x+i_y}\phi_i(\tau) \gamma^T_i \kappa^y l^z\sigma^z \gamma_i)}.
\end{equation}
The superscripts $\left\{0,x,y,z\right\}$ in $\kappa,l,\sigma$ label U(2) group generators (i.e., identity together with three Pauli matrices) for Majorana, layer and spin degrees of freedom, inverse temperature $\beta=L_\tau\Delta\tau$, $i,j$ indicate site positions, and summation over inner degrees of freedom is implied. The first term in the exponent is nothing but a fourfold degenerate tight-binding Hamiltonian which in Fock basis is
\begin{equation}
-\Delta\tau H_0 = \Delta\tau\sum_{i,j} t_{i,j} (-a^{\dagger}_i l^0 \sigma^0 a_j + a_i l^0 \sigma^0 a_j^{\dagger}).
\end{equation}
The interaction of Majorana fermions is involved through the local coupling of the scalar field $\phi$ in the second term on the exponent which is $-2\Delta\tau g \sum_{i}(-1)^{i_x+i_y}\phi_i(\tau) a^{\dagger}_i l^z\sigma^z a_i$ in Fock basis, where the order of $\phi$ drives the anti-ferromagnetic order for layer-spin polarization on a square lattice. One obvious anti-unitary symmetry for such a system is $\mathcal{T}=-\mathds{1} \otimes \mathrm{i} \kappa^x l^x\sigma^y \mathcal{K}$, which corresponds to the spatially homogeneous superconducting order whose local order parameter is $\gamma_i^T \kappa^x l^x\sigma^y \gamma_i=\mathrm{i}(a_i l^x\sigma^y a_i - a_i^\dagger l^x\sigma^y a_i^\dagger)$. By definition, the expectation value of the two-point correlation function for such FGS, e.g. $\left\langle \gamma_i^T(\tau) \kappa^y l^z\sigma^z \gamma_j(\tau') \right\rangle$, is
\begin{equation}
\left\langle \gamma_i^T(\tau) \kappa^y l^z\sigma^z \gamma_j(\tau') \right\rangle = \frac{\int [\mathcal{D} \phi] e^{-S(\left\{\phi\right\})} \left\langle \gamma_i^T(\tau) \kappa^y l^z\sigma^z \gamma_j(\tau') \right\rangle_{\phi}}{\int [\mathcal{D} \phi'] e^{-S(\left\{\phi'\right\})}}.
\end{equation}
With the Gaussian approximation of the real action $S(\left\{\phi\right\})$ around its critical saddle point, the generated configurations of $\phi$ are determined by the correlation as we engineered in the previous section, i.e.
\begin{equation}
\langle \phi(r) \phi(0) \rangle \equiv \frac{\int [\mathcal{D} \phi] e^{-S(\left\{\phi\right\})} \phi(r) \phi(0)}{\int [\mathcal{D} \phi'] e^{-S(\left\{\phi'\right\})}}=\frac{1}{(r/r_0+1)^{\Delta}},
\end{equation}
For fixed $\Delta,z$, we can generate a group of critical configurations $\left\{\phi\right\}$ and measure bilinear correlations of fermions based on that.

\begin{figure}[htp]
\centering
\includegraphics[width=1.0\columnwidth]{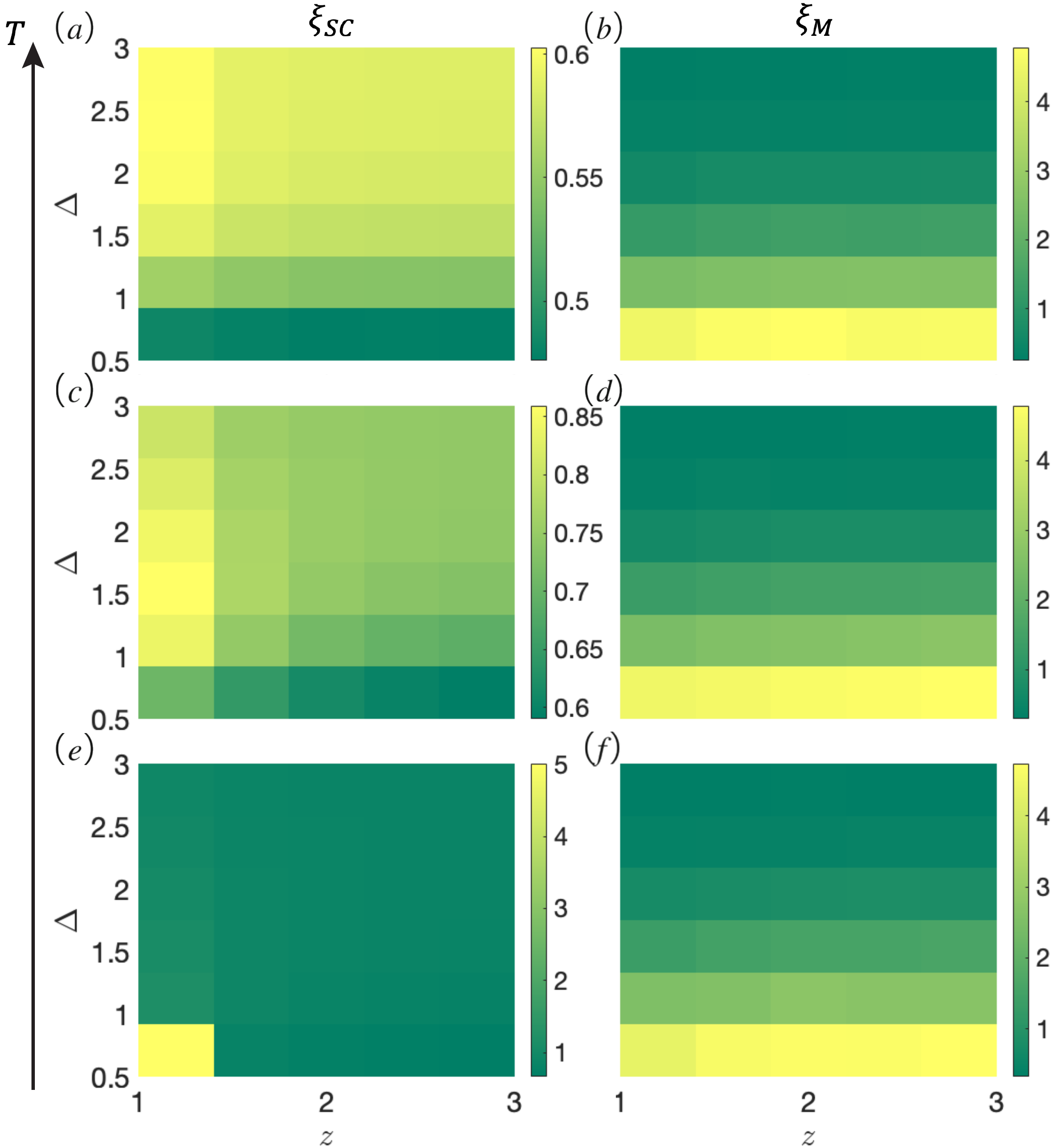}
\caption{Competition between superconducting and anti-ferromagnetic orders from correlation length $\xi_{SC}$ in (a,c,e) and $\xi_{M}$ in (b,d,f) as a function of critical exponents $\Delta,z$. The imaginary-time length $\beta$ is $0.1L,0.2L,0.3L$ for (a,b),(c,d),(e,f) where system size $L=12$.}
\label{SCZZ}
\end{figure}

By scanning $\Delta\in[0.5,3],z\in[1,3]$, one can observe the competition between anti-ferromagnetic order and superconducting order selected by the stable saddle point of $S(\left\{\phi\right\})$ and the anti-unitary symmetry respectively as shown in Fig.~\ref{SCZZ}. In the simulation, we choose $L\times L$ square lattice tight-binding model with the nearest, next-nearest-neighbor hopping $t=-3$, $t'=-0.35t$ and chemical potential $\mu=-3.5$ for doping holes. The coupling strength $g=5$ and imaginary time direction is discretized with $\Delta\tau=0.05$. We fix $L=12$ and tune temperature by $\beta=0.1L,0.2L,0.3L$. In this way, we will not do finite-size scaling and only focus on finite-temperature instability at different critical exponents $\Delta,z$ by defining the correlation length $\xi_{SC},\xi_{M}$ as
\begin{equation}
    \label{cl}
    \xi=\frac{1}{\Delta q}\sqrt{\frac{\chi(Q)}{\chi(Q+\Delta q)}-1},
\end{equation}
where $\chi(q)\equiv \mathcal{F}[\left\langle M(r)M(0) \right\rangle]$ is the static structure factor peaking at momentum $Q$ from the spatial Fourier transformation of bilinear correlation $\left\langle M(r)M(0) \right\rangle$ with local-order measurement operator $M(r)$, and $\Delta q$ is the smallest momentum $2\pi/L$ for size $L$. With this definition, the equal-time correlations for superconducting and anti-ferromagnetic at $i,j$ are $\left\langle a_i l^x\sigma^y a_i a^\dagger_j l^x\sigma^y a^\dagger_j + a^\dagger_i l^x\sigma^y a^\dagger_i a_j l^x\sigma^y a_j \right\rangle$ and $4\left\langle a^{\dagger}_i l^z\sigma^z a_i a^{\dagger}_j l^z\sigma^z a_j \right\rangle$.

From Fig.~\ref{SCZZ} (a,c,e), we can conclude that the superconducting instability is more sensitive to the fluctuation $\phi$ along imaginary time direction. A larger $z$ or smaller $\beta$ suppressing the quantum fluctuation will eliminate superconductivity. Above the transition temperature, The small-/large-$\Delta$ limit corresponds to order/disorder stable fixed point and $\xi_{SC}$ should have the maximal value in between, as shown in (a,c) in Fig.~\ref{SCZZ}. In contrast, $\xi_{M}$ is less sensitive to quantum fluctuation and almost only relies on spatial correlation exponent $\Delta$ as shown in Fig.~\ref{SCZZ} (b,d,f). Nevertheless, the superconducting instability at small $z$ weakens its correlation length, indicating the competition between such two mechanisms for spontaneous symmetry breaking.

\section{Simulation of bosonic FGS with engineered critical fluctuations}\label{s7}
In this section, we will explore an approximate bosonic FGS example. For simplicity, we only keep the position degrees of freedom for phase-space boson such that the FGS realizes a 2+1D classical model, where a traceless-symmetric matrix $L$ in $\mathcal{T}=L\mathcal{K}$ corresponds to a bilinear nematic order $\bar{\alpha}^T L \bar{\alpha}$. Different from fermions, bosons can condense by themselves (i.e., $\left\lvert G_{ij}\right\rvert^2$ becomes finite for separated $i,j$) and its bilinear will be ordered automatically even at a stable saddle point as discussed in the previous section. To rule out such a case, we couple the boson with a trivial gauge field (i.e., a gauge field without any dynamics) so that a single-boson operator is gauge dependent and can never be ordered according to Elitzur's theorem~\cite{Elitzur1975impossibility}. Since FGS require no higher order terms other than bilinear, one also needs to introduce a constraint term to control the magnitude and avoid divergence.

We use an effective $S^3$ rotor model by taking the saddle-point approximation of the constraint field to realize requirements above. The partition function for fixed $\phi$ is
\begin{eqnarray}
    Z_{\phi}&\equiv&\int [\mathcal{D} \bn] \sum_{\left\{s_{ij}=\pm1\right\} } e^{-H[\phi,s,\bn]} \\
    &=&\int [\mathcal{D} \bn] \sum_{\left\{s_{ij}=\pm1\right\} } \frac{1}{2^{N_s}} e^{\sum_{\left\langle i,j\right\rangle }J s_{ij} \bn_i \cdot \bn_j + \sum_i g \phi_i \bn_i^{T} l^z \sigma^z \bn_i}. \nonumber
\end{eqnarray}
Here $\bn$ with $\bn^2=1$ is a four-component rotor on $S^3$ sphere, which can be realized from four-component boson field $\bq$ by $\int d \lambda \int d \bq e^{\mathrm{i}\lambda(\bq^2-1)}\cdots = \int d \bn\cdots$. $\left\langle i,j\right\rangle$ indicates the nearest neighbor coupling along three directions of a cubic lattice, $J,g$ are positive constants and $s_{ij}=\pm1$ is a $\mathbb{Z}_2$ gauge field defined on each bond with the total amount $N_s$. Since the system has a local gauge symmetry that $\bn_i\rightarrow-\bn_i,s_{ij}\rightarrow-s_{ij}$ where $s_{ij}$ runs over all the connected bonds with $i$, the gauge dependent order $\left\langle \bn_i\right\rangle\neq 0$ is prevented and we expect enlarging $J$ will drive the system into a nematic phase. The measurement of two-point correlation function, e.g. $\left\langle \bn_a^T l^z\sigma^z \bn_b \right\rangle$, is
\begin{equation}
    \label{mc}
    \left\langle \bn^T_a l^z\sigma^z \bn_b \right\rangle = \frac{\int [\mathcal{D} \phi] e^{-S(\left\{\phi\right\})} \left\langle \bn_a^T l^z\sigma^z \bn_b \right\rangle_{\phi}}{\int [\mathcal{D} \phi'] e^{-S(\left\{\phi'\right\})}},
\end{equation}
where $\left\langle \bn_a^T l^z\sigma^z \bn_b \right\rangle_{\phi}\equiv\frac{\int [\mathcal{D} \bn] \sum_{\left\{s_{ij}\right\} } e^{-H[\phi,s,\bn]} \bn_a^T l^z\sigma^z \bn_b}{\int [\mathcal{D} \bn'] \sum_{\left\{s'_{ij}\right\} } e^{-H[\phi,s',\bn']}} = \frac{\int [\mathcal{D} \bn] \prod_{\left\langle i,j\right\rangle } \cosh(J \bn_i \cdot \bn_j) e^{\sum_i g \phi_i \bn_i^{T} l^z \sigma^z \bn_i} \bn^T_a l^z\sigma^z \bn_b}{\int [\mathcal{D} \bn'] \prod_{\left\langle i,j\right\rangle } \cosh(J \bn'_i \cdot \bn'_j) e^{\sum_i g \phi_i (\bn'_i)^T l^z \sigma^z \bn'_i}}$. Note this is not a measurement of a Gaussian state yet, but one will see the whole two-point correlation function can be seen approximately as a measurement of FGS since
\begin{eqnarray}
    &&\left\langle \bn^T_a l^z\sigma^z \bn_b \right\rangle \\
    &=& \int [\mathcal{D} \phi] \sum_{\left\{s_{ij}=\pm1\right\}} p(\left\{\phi,s_{ij}\right\}) \frac{\int [\mathcal{D} \bn] e^{-H[\phi,s,\bn]} \bn^T_a l^z\sigma^z \bn_b}{\int [\mathcal{D} \bn'] e^{-H[\phi,s,\bn']}}, \nonumber
\end{eqnarray}
where the positive-definite probability density function $p(\left\{\phi,s_{ij}\right\})\equiv\frac{e^{-S(\left\{\phi\right\})}\int [\mathcal{D} \bn] e^{-H[\phi,s,\bn]}}{\int [\mathcal{D} \phi'] e^{-S(\left\{\phi'\right\})}\int [\mathcal{D} \bn'] \sum_{\left\{s'_{ij}\right\} } e^{-H[\phi,s',\bn']}}$ corresponds to $\frac{e^{-S(\left\{\phi\right\})}}{\int [\mathcal{D} \phi'] e^{-S(\left\{\phi'\right\})}}$ in FGS by replacing $\left\{\phi\right\}$ with $\left\{\phi,s_{ij}\right\}$. Now the measurement is Gaussian under saddle-point approximation for $\lambda$ in $\int d \bn\cdots = \int d \lambda \int d \bq e^{\mathrm{i}\lambda(\bq^2-1)}\cdots$, which is reasonable only when the dimension of local rotor is large enough. We will numerically confirm the qualitative estimation works well even when the dimension is 4 in this example.

In practical simulation, we generate critical configuration $\left\{\phi\right\}$ from the method mentioned in previous section for fixed $z=1$ and tune $\Delta\in[0.5,3]$, and use Eq.~\eqref{mc} to compute observables from Monte Carlo sampling $\bn$. The approximate FGS has anti-unitary symmetries $\mathcal{T}=L\mathcal{K}$, where $L=\mathds{1} \otimes l^0\sigma^z,\mathds{1} \otimes l^z\sigma^0,\mathds{1} \otimes l^x\sigma^x,\mathds{1} \otimes l^y\sigma^y$ are real orthogonal traceless-symmetric matrices. One can conclude that within all directions of four-dimensional nematic orders, $\bn_i^{T} L \bn_i$ are selected to be ordered departing from stable saddle point of $\bn_i^{T} l^z\sigma^z \bn_i$ order.

\begin{figure}[htp]
\centering
\includegraphics[width=1.0\columnwidth]{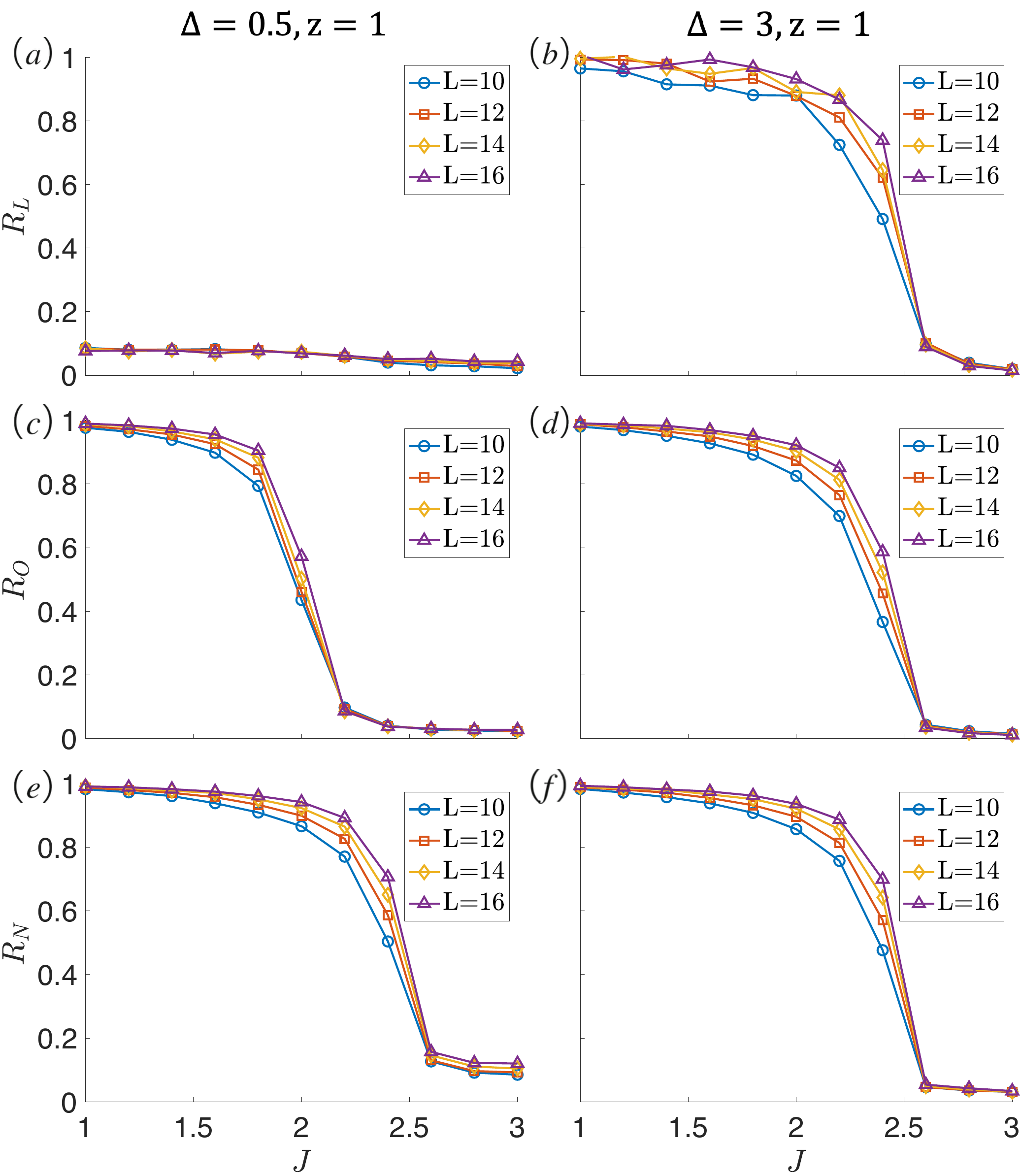}
\caption{Correlation ratio for different order parameters $\bn_i^{T} l^z\sigma^z \bn_i$ in (a,b), $\bn_i^{T} l^0\sigma^z \bn_i$ in (c,d) and $\bn_i^{T} l^0\sigma^x \bn_i$ in (e,f). (a,c,e) are results at critical exponents $\Delta=0.5,z=1$ and (b,d,f) are results at $\Delta=3,z=1$. The three dimensions are scaled together with system size $L=10,12,14,16$.}
\label{BALL}
\end{figure}

The simulation result indeed shows such a case. We set $g=1$, tune $J\in[1,3]$, scale three directions of the cubic lattice together and measure the correlation ratio defined as $R\equiv \frac{S(Q+\Delta q)}{S(Q)}$ where $S(Q)\equiv\mathcal{F}[\left\langle M(r)M(0) \right\rangle]$ is the dynamic structure factor with three-dimensional Fourier transformation of bilinear correlation $\left\langle M(r)M(0) \right\rangle$. Compared with Eq.~\eqref{cl}, one can see the parameter $R$ as a normalized correlation length, where a smaller value corresponds to a longer correlation. In Fig.~\ref{BALL}, we use $R_L,R_O,R_N$ to represent correlation ratios for $\bn_i^{T} l^z\sigma^z \bn_i,\bn_i^{T} l^0\sigma^z \bn_i,\bn_i^{T} l^0\sigma^x \bn_i$ orders. As shown in Fig.~\ref{BALL} (a,b), the order $\bn_i^{T} l^z\sigma^z \bn_i$ coupled with $\phi$ field follows the behavior of $\phi$ closely at small $J$ and a larger $J$ will break such a coupling and drive the system into nematic order. Compared with $\bn_i^{T} l^0\sigma^x \bn_i$ in Fig.~\ref{BALL} (e,f), the order parameter $\bn_i^{T} l^0\sigma^z \bn_i$ selected by the anti-unitary symmetry in Fig.~\ref{BALL} (c,d) develops a long correlation length earlier at $\Delta=0.5$ where the fluctuation of $\phi$ is large. In the Monte Carlo update of $\bn$, we combine local update with Wolff cluster update~\cite{wolff1989collective} to reduce the autocorrelation time. For building up a Wolff cluster, we have used two O(2) symmetries between components 1,4 and 2,3 in vector $\bn$. The acceptance probability for bond $ij$ is
\begin{equation}
    P_{ij}=1-e^{\min\left\{0,\log[\cosh(J\bn_i\cdot\bn_j)]-\log[\cosh(J \mathcal{R}[\bn_i]\cdot\bn_j)]\right\} }.
\end{equation}
Here $\mathcal{R}[\bn_i]$ is the mirror reflection of vector $\bn_i$ with mirror plane defined by 1,4 or 2,3 components in $\bn$.

\section{Represent FGS with PEPS ansatz}\label{s8}
In this section, we discuss the possibility of representing FGS with PEPS ansatz. A general density matrix with mutual information satisfying area law can be represented with a PEPS after purification~\cite{Wolf2008area,cirac2021matrix}, though a Fermi liquid pure state can easily violate the mutual information requirement. However, when anti-unitary symmetry $\mathcal{T}=L\mathcal{K}$ exists for fermionic FGS, one could derive the Fermi liquid solution is unstable if the orthogonal matrix $L$ is block diagonal in momentum space so that the anti-unitary symmetry preserves after renormalization~\cite{grossmanrobust2023}. This result naturally brings out a question: can PEPS ansatz describe fermionic FGS with anti-unitary symmetries?

We partially answer this question with a straightforward construction representing every Gaussian state by a Gaussian PEPS with bond dimension $D$~\cite{mortier2022tensor,mortier2023finite}. If Monte Carlo can sample FGS with $N_s$ samples, then the bond dimension of PEPS ansatz for such FGS is upper bounded by $N_s D$. We argue this construction can work well in the saddle point phase, but will fail in the non-local contraction phase, where the maximal connected correlation exceeds the expressivity of the PEPS ansatz in this construction. Below, we will first take every Gaussian state is \emph{Hermitian} condition and give a rigorous upper bound of the expressivity via conditional mutual information, and then argue this bound should preserve when anti-unitary symmetry exists while the Hermitian requirement is released.

To start, we separate the whole system into two identical regions $A$ and $\bar{A}$. For a fixed $\phi$, the entanglement from the reduced density matrix of a Gaussian state is purely determined by equal-time Green's function
\begin{equation}
    G=\left(\begin{matrix}G_A & G_{O} \\ - G^{T}_{O} & G_{\bar{A}} \end{matrix}\right).
\end{equation}
If the Gaussian state is Hermitian, one can define a conditional mutual information~\cite{Groisman2005quantum}
\begin{eqnarray}
    \label{iaa}
    I(A;\bar{A}|\phi)&\equiv&\Tr[\rho_\phi\log(\rho_\phi)-\rho_A\otimes\rho_{\bar{A}}\log(\rho_A\otimes\rho_{\bar{A}})] \nonumber\\
    &=&\Tr[G\log(G)-G_{A}\oplus G_{\bar{A}}\log(G_{A}\oplus G_{\bar{A}})] \nonumber\\
    &\geqslant& \Tr[(G-G_{A}\oplus G_{\bar{A}})^2] \nonumber\\
    &=&2\left\lVert G_O\right\rVert_F^2,
\end{eqnarray}
where $\rho_A\equiv \Tr_{\bar{A}}(\rho_\phi),\rho_{\bar{A}}\equiv \Tr_{A}(\rho_\phi)$ are reduced density matrices, and von Neumann entropy in our definition of Majorana basis (i.e., $\gamma^2=\frac{1}{2}$) is $S_\phi\equiv -\Tr[\rho_\phi\log(\rho_\phi)] = -\Tr[G\log(G)]$ (see Appendix.~\ref{ab} for detailed derivation). Since eigenvalues of the Green's function matrix are paired with the summation equal to 1, we have $\Tr[G\log(G)]=\Tr[(\mathds{1}-G)\log(\mathds{1}-G)]$, and the lower bound in Eq.~\eqref{iaa} is derived following the proof of the Pinsker's inequality
\begin{eqnarray}
    &&\Tr[G\log(G)-G_{A}\oplus G_{\bar{A}}\log(G_{A}\oplus G_{\bar{A}})] \nonumber\\
    &=& \frac{1}{2} \{ \Tr[G\log(G)-G\log(G_{A}\oplus G_{\bar{A}})] \nonumber\\
    &+& \Tr[G\rightarrow\mathds{1}-G, G_{A}\oplus G_{\bar{A}}\rightarrow\mathds{1}-G_{A}\oplus G_{\bar{A}}]\} \nonumber\\
    &=& \frac{1}{2} \sum_{i,j}\left\lvert \left\langle i|j\right\rangle \right\rvert^2 \left[ g_i \log(\frac{g_i}{g'_j}) + (1-g_i) \log(\frac{1-g_i}{1-g'_j}) \right] \nonumber\\
    &\geqslant& \sum_{i,j}\left\lvert \left\langle i|j\right\rangle \right\rvert^2 (g_i-g'_j)^2 \nonumber\\
    &=& \Tr[(G-G_{A}\oplus G_{\bar{A}})^2],
\end{eqnarray}
where $i,j$ label eigenstates of $G,G_{A}\oplus G_{\bar{A}}$, and $g_i,g'_j$ are corresponding eigenvalues.

From Ref.~\cite{Wolf2008area}, one has the inequality between PEPS ansatz bond dimension $D_{\phi}$ and $\left\lVert G_O\right\rVert_F^2$ via mutual information
\begin{equation}
    \left\lvert \partial A\right\rvert \log(D_{\phi}) \geq \frac{1}{2}I(A;\bar{A}|\phi) \geq \left\lVert G_O\right\rVert_F^2,
\end{equation}
where $\left\lvert \partial A\right\rvert$ is the boundary length between subsystems $A$ and $\bar{A}$. Thus the average bond dimension should satisfy
\begin{equation}
    \label{dg}
    \left\lvert \partial A\right\rvert \log(\left\langle D_\phi\right\rangle) \geq \left\lvert \partial A\right\rvert \left\langle\log( D_\phi) \right\rangle \geq \left\langle \left\lVert G_O\right\rVert_F^2 \right\rangle.
\end{equation}

This is the rigorous bound derivation for Hermitian Gaussian state. We find the maximal connected correlation function of bilinear order parameters between two regions shares the same bound when anti-unitary symmetry exists and the Hermitian requirement can be absent. Assuming the real orthogonal matrix $L$ in the anti-unitary symmetry is block diagonal as
\begin{equation}
    L=\left(\begin{matrix}L_A & 0 \\ 0 & L_{\bar{A}} \end{matrix}\right).
\end{equation}
From Eq.~\eqref{gf} the Green's function off-diagonal block has the relation
\begin{equation}
    L_A G_{O} L_{\bar{A}}^T=G_O^*.
\end{equation}
And similar with the derivation of Eq.~\eqref{CF}, the connected correlation function between regions $A$ and $\bar{A}$ for a Gaussian state has
\begin{eqnarray}
    \label{cll}
    &&\left\langle M_{\bar{A}} \otimes M_{A} \right\rangle_\phi-\left\langle M_{\bar{A}} \right\rangle_\phi\left\langle M_{A} \right\rangle_\phi \nonumber\\
    &=&2\Tr[O_{\bar{A}}^T T^g O_{A} G_O] \Tr[L_{\bar{A}}^T T^g L_A G_O]^* \nonumber\\
    &\leq&2\left\lVert G_O\right\rVert_F^2,
\end{eqnarray}
where $O_A,O_{\bar{A}}$ are real orthogonal matrices sharing the same dimension in order parameters $M_A\equiv\gamma_{A}^T\mathrm{i}O_A\gamma_{A},M_{\bar{A}}\equiv\gamma_{\bar{A}}^T\mathrm{i}O_{\bar{A}}\gamma_{\bar{A}}$, and the equality holds iff $O_A=L_A,O_{\bar{A}}=L_{\bar{A}}$. Since the bond dimension should count total amount of connected correlations between two systems~\cite{Wolf2008area,Groisman2005quantum}, we expect the similar formula $\left\lvert \partial A\right\rvert \log(\left\langle D_\phi\right\rangle) \gtrsim \left\langle \left\lVert G_O\right\rVert_F^2 \right\rangle$ holds in a general fermionic FGS with Kramers anti-unitary symmetry.

From such a formula one can tell, if only the boundary contributes the connected correlation, which is the stable saddle point case, $\left\lVert G_O\right\rVert_F^2\sim\left\lvert \partial A\right\rvert$ and finite $D$ satisfies this inequality. However, when the system enters the non-local contraction phase, i.e. $\left\langle\left\lVert G_O\right\rVert_F^2\right\rangle\sim\mathcal{O}(N^2)$, the average bond dimension should grow exponentially which makes this specific construction impractical.

\section{Conclusion and discussion}\label{s9}
Inspired by the anti-unitary symmetry in sign-problem-free fermionic QMC~\cite{zhang2025constraints,zhang2025bilayer}, we develop the concept of FGS for both fermions and bosons. The anti-symmetric (traceless-symmetric) real orthogonal matrix $L$ in $\mathcal{T}=L\mathcal{K}$ for anti-unitary symmetric fermionic (bosonic) FGS determines the maximal bilinear order parameter correlation of the non-local contraction term. The stability of the fluctuation saddle point dominates the relative magnitude of direct contraction and non-local contraction terms, demonstrating the competition between two possible spontaneous symmetry breaking phases. Since a stable Fermi liquid is ruled out for fermionic FGS~\cite{grossmanrobust2023}, we also discuss the possibility of representing fermionic FGS with PEPS ansatz. A straightforward construction that builds one PEPS for one Gaussian state satisfies the entanglement requirement in the saddle point phase, but will be impractical in the non-local contraction phase, though a more general construction is still possible there.

One should also note, our definition of FGS does not necessarily require the imaginary time path and the result of correlations naturally applies to a general Keldysh formalism. Thus the anti-unitary symmetries of FGS may also be relevant with observed exact steady states in non-equilibrium systems with hidden time-reversal symmetry~\cite{roberts2021hidden,lingenfelter2026exact}.

In nature, bilinear orders such as magnetism, charge density wave and superconductivity for fermions and squeezing and nematic order for bosons are commonly seen. Some of them are generally robust against perturbation while some are very fragile. Our study suggests a possible way of stabilizing the target bilinear order, i.e. if the order can be seen as emerging from free fermions (bosons) interacting with a fluctuating environment with a real action, then prohibit the coupling that breaks the corresponding anti-unitary symmetry will protect the target bilinear order phase.

\begin{acknowledgements}
I sincerely thank Nick Bultinck, Bram Vancraeynest-De Cuiper and Jutho Haegeman for sharing their knowledge on model design, Cauchy-Schwarz inequality and symplectic property, and for the help in polishing the manuscript. I thank Jiangping Hu, Congjun Wu for helpful discussion on related topics. This research was supported by the European Research Council under the European Union Horizon 2020 Research and Innovation Programme via Grant Agreement No. 101076597-SIESS. The computational resources and services used in this work were partially provided by the VSC (Flemish Supercomputer Center), funded by the Research Foundation
- Flanders (FWO) and the Flemish Government. I acknowledge EuroHPC Joint Undertaking for awarding me access to MareNostrum5 hosted by Barcelona Supercomputing Center, Spain.
\end{acknowledgements}

\bibliography{FGS}

\begin{thebibliography}{44}
\bibitem [{ant()}]{antiunitary}%
  \BibitemOpen
  \href@noop {} {\bibinfo  {journal} {This definition of anti-unitary symmetry is based on matrix algebra $L^TQ_\phi^*L=Q_\phi$ where $Q_\phi$ is the matrix within bilinears of $B_\phi(\tau)$. The real orthogonal (symplectic or anti-symplectic) requirement of $L$ for fermions (bosons) will be used in later linear/anti-linear canonical basis transformation}\ }\BibitemShut {NoStop}%
\end{thebibliography}

\appendix
\section{Canonical basis transformation}\label{ct}
We derive the canonical basis transformation for Majorana fermions (phase-space bosons) in this section, which can be linear or anti-linear. Under linear transformation $\bar{c}\rightarrow L\bar{c}$, the canonical anti-commutation (commutation) relation becomes
\begin{eqnarray}
    \label{cr}
    \delta_{ac}=\left\{\gamma_a',\gamma_c'\right\}&=&\left\{L_{ab}\gamma_b,L_{cd}\gamma_d\right\}=L_{ab}L_{cb} \\
    \mathrm{i}\Omega_{ac}=\left[\bar{\alpha}_a',\bar{\alpha}_c'\right]&=&\left[L_{ab}\bar{\alpha}_b,L_{cd}\bar{\alpha}_d\right]=L_{ab}\mathrm{i}\Omega_{bd}L_{cd}, \nonumber
\end{eqnarray}
where symplectic matrix
\begin{equation}
    \Omega=\left(\begin{matrix}0 & \mathds{1} \\ -\mathds{1} & 0 \end{matrix}\right).
\end{equation}
This together with the Hermitian requirement $L_{ab}\bar{c}_b=L_{ab}^*\bar{c}_b$ implies the linear transformation matrix $L$ should be real orthogonal (symplectic) for Majorana fermions (phase-space bosons). Since the Hilbert space trace is invariant after linear canonical basis transformation, we have
\begin{equation}
    \Tr(\rho \bar{c}_{a} \bar{c}_c)=\Tr(\rho(\bar{c}\rightarrow L\bar{c}) L_{ab}\bar{c}_{b} L_{cd}\bar{c}_d).
\end{equation}
The requirement of anti-linear canonical basis transformation on $L$ is the same for Majorana fermions, while the conjugation contributes an additional minus sign for phase-space boson formula in Eq.~\eqref{cr}. Thus $L$ should be real orthogonal (anti-symplectic) for Majorana fermions (phase-space bosons). The Hilbert space trace has an additional conjugation after anti-linear canonical basis transformation, i.e.
\begin{equation}
    \Tr(\rho \bar{c}_{a} \bar{c}_c)=\Tr(\rho^*(\bar{c}\rightarrow L\bar{c}) L_{ab}\bar{c}_{b} L_{cd}\bar{c}_d)^*.
\end{equation}

\section{Fermi surface of fermionic FGS}
\begin{figure}[htp]
\centering
\includegraphics[width=1.0\columnwidth]{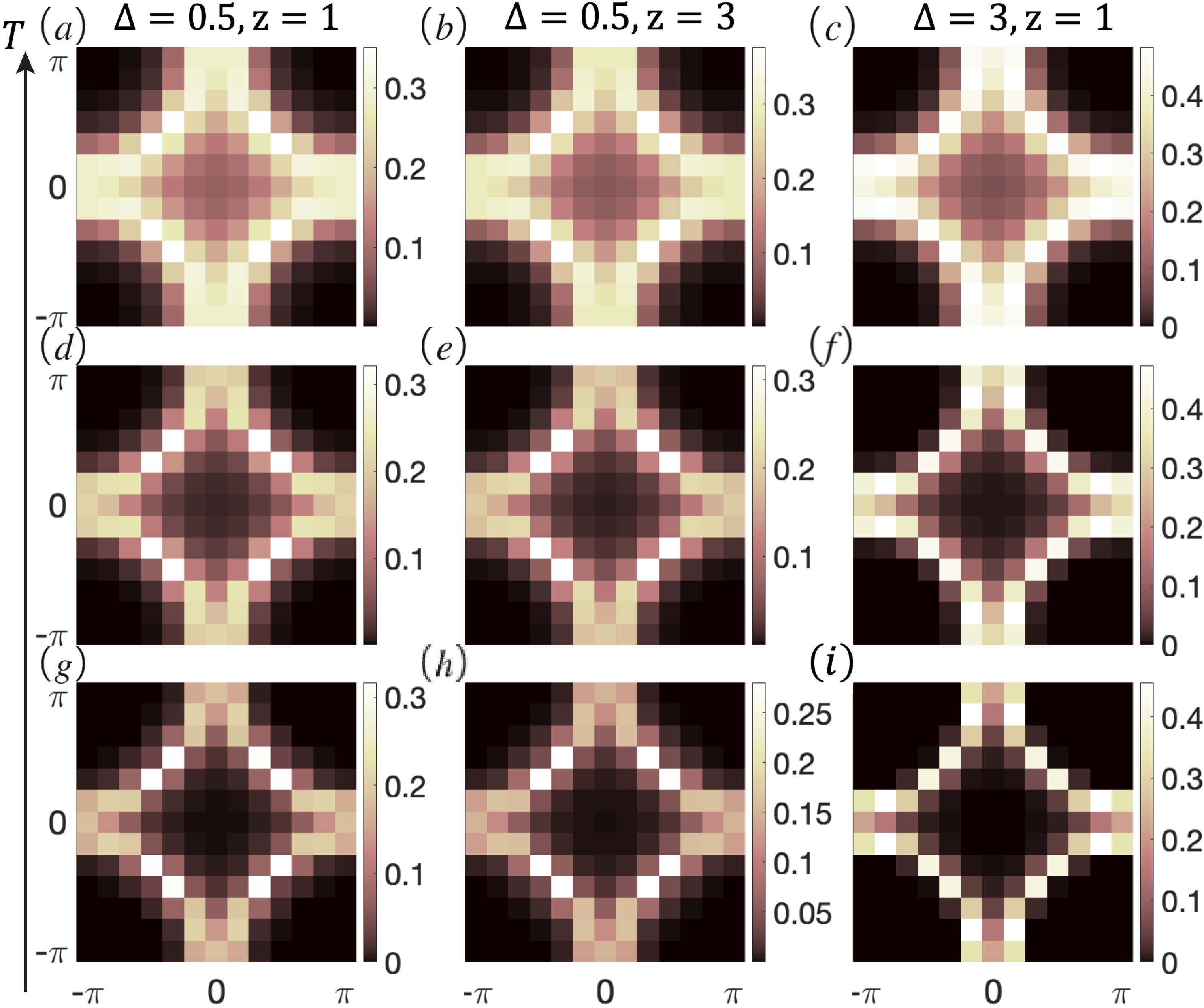}
\caption{Fermi surface indicated from $G(\beta/2,k)$ at different temperature and critical exponents $\Delta,z$. The imaginary-time length $\beta$ is $0.1L,0.2L,0.3L$ for (a,b,c),(d,e,f),(g,h,i) where system size $L=12$.}
\label{FSall}
\end{figure}

Another interest for fermionic system is the Fermi surface. Luttinger's theorem estimates the volume enclosed by the Fermi surface is determined only by the particle density~\cite{Oshikawa2000,Oshikawa2000_2,Hastings2004,FLstar}, which may still hold across the critical region of a non-Fermi liquid whose Fermi surface is defined according to the poles of real-frequency Green's function $\Re[G^{-1}(0,k_F)]=0$ when the quasi-particle residue is zero. As shown in Fig.~\ref{FSall}, we take the fermionic FGS described in the main text and use imaginary-time Green's function at $\beta/2$ to approximate zero-frequency spectral function at low temperature, i.e.
\begin{eqnarray}
    G(\beta/2,k)&=&\int d\omega \frac{A(\omega,k)}{2\cosh(\beta\omega/2)} \\
    &\approx& \frac{Z(k)}{2\cosh(\beta\varepsilon_k /2)}  + \frac{\pi}{\beta} A_c(\omega=0,k), \nonumber
\end{eqnarray}
where $A_c(\omega=0,k)$ is the continuum part of spectral function $A(\omega,k)$ at zero-frequency. For a Fermi liquid with quasi-particle residue $Z(k)=1$, $G(\beta/2,k_F)=0.5$ on the Fermi surface at low temperature limit. Three groups of critical exponents $(\Delta,z)=(0.5,1),(0.5,3),(3,1)$ are used for each column representing different space-time fluctuation of $\phi$ field. One can see the low-frequency spectra are affected most relevantly with the anti-ferromagnetic correlation length, which is sensitive to $\Delta$ but not $z$, thus a larger $z$ can cause the Fermi arc away from superconduct region. Even though the suppression of quasi-particle residue is momentum dependent, one should notice the Fermi surface has the same shape which indicates Luttinger's theorem is not violated for FGS across the critical region with different critical exponents.

\section{Derivation of von Neumann entropy}\label{ab}
In this section, we prove with the convention $\gamma^2=\frac{1}{2}$, von Neumann entropy of a fermionic Gaussian state can be expressed with Green's function as
\begin{equation}
    S\equiv -\Tr[\rho\log(\rho)] = -\Tr[G\log(G)],
\end{equation}
where $G_{ij}\equiv\Tr(\rho\gamma_{j}\gamma_{i})$.

According to Ref.~\cite{lucas2021bosonic} where the convention $\gamma'^2=1$ is taken, the correlation matrix $\Gamma_{ij}\equiv\frac{\mathrm{i}}{2}\Tr(\rho[\gamma'_i,\gamma'_j])=\mathrm{i}\Tr(\rho[\gamma_i,\gamma_j])=\mathrm{i}(\delta_{ij}-2G_{ij})$. The anti-symmetric $\Gamma$ matrix can be brought in standard form together with $G$ matrix with the same orthogonal canonical transformation
\begin{eqnarray}
    \Gamma&=&O \bigoplus_{i=1}^N \left(\begin{matrix}  0 & \lambda_i\\ -\lambda_i & 0 \end{matrix} \right) O^T \nonumber\\
    G&=&O \bigoplus_{i=1}^N \left(\begin{matrix}  \frac{1}{2} & \frac{\mathrm{i}\lambda_i}{2}\\ -\frac{\mathrm{i}\lambda_i}{2} & \frac{1}{2} \end{matrix} \right) O^T.
\end{eqnarray}
The most familiar von Neumann entropy formula based on $\lambda$ of $\Gamma$ matrix is
\begin{equation}
    S=-\sum_{i=1}^N\left[\frac{1+\lambda_i}{2}\log(\frac{1+\lambda_i}{2})+\frac{1-\lambda_i}{2}\log(\frac{1-\lambda_i}{2})\right].
\end{equation}
Since $\frac{1\pm\lambda_i}{2}$ corresponds to two eigenvalues of Green's function $G$ matrix in that $2\times2$ block, one can derive the formula we use in the main text that
\begin{equation}
    S= -\Tr[G\log(G)].
\end{equation}

\end{document}